\documentclass[11pt,preprint]{aastex}
\usepackage{graphicx}

\def\lya{Lyman-$\alpha$}

\def\lstar{\ifmmode {L_\star}\else
                ${L_\star}$\fi}
\def\phistar{\ifmmode {\phi_\star}\else
                ${\phi_\star}$\fi}
\def\ergcm2s{\ifmmode {\rm\,erg\,cm^{-2}\,s^{-1}}\else
                ${\rm\,ergs\,cm^{-2}\,s^{-1}}$\fi}
\def\ergsec{\ifmmode {\rm\,erg\,s^{-1}}\else
                ${\rm\,ergs\,s^{-1}}$\fi}
\def\kmsMpc{\ifmmode {\rm\,km\,s^{-1}\,Mpc^{-1}}\else
                ${\rm\,km\,s^{-1}\,Mpc^{-1}}$\fi}

\def\sun{\odot}

\begin{document}

\title{Sizing up Lyman-alpha and Lyman Break Galaxies}

\author{Sangeeta Malhotra\altaffilmark{1},  
James E. Rhoads\altaffilmark{1}, 
Steven L.  Finkelstein\altaffilmark{2}, 
Nimish Hathi\altaffilmark{3}, 
Kim Nilsson\altaffilmark{4}, 
Emily McLinden\altaffilmark{1}, and 
Norbert Pirzkal\altaffilmark{5}}

\altaffiltext{1}{School of Earth and Space Exploration, Arizona State University}
\altaffiltext{2}{George P. and Cynthia Woods Mitchell Institute for Fundamental Physics and Astronomy, Department of Physics and Astronomy, Texas A\&M University}
\altaffiltext{3}{Carnegie Observatories, Pasadena, CA}
\altaffiltext{4}{ST-ECF, Garching}
\altaffiltext{5}{Space Telescope Science Institute}

\begin{abstract}
  We show that populations of high redshift galaxies grow more
  luminous as they grow in linear size. This is because the brightness
  per unit area has a distinct upper limit due to the self-regulating
  nature of star-formation.  As a corollary, we show that the observed
  increase in characteristic luminosity of Lyman Break Galaxies
  (L$_*$) with time can be explained by their increase in size, which
  scales as $H(z)^{-1}$.  In contrast, \lya\ selected galaxies have a
  characteristic, constant, small size between redshift $z=2.25$ and
  6.5.  Coupled with a characteristic surface brightness, this can
  explain their non-evolving ultraviolet continuum luminosity
  function. This compact physical size seems to be a critical
  determining factor in whether a galaxy will show \lya\ emission.  We
  base these conclusions on new size measurements for a sample of
  about 170 \lya\ selected galaxies with {\it Hubble Space Telescope}
  broad band imaging, over the redshift range $2.25 < z < 6$.  We
  combine these with a similar number of \lya\ selected galaxies with
  half-light radii from the literature, and derive surface
  brightnesses for the entire combined sample.
\end{abstract}

\keywords{galaxies: high-redshift}

\section{Introduction}

The observed properties of high redshift galaxies provide crucial
constraints on galaxy formation scenarios.  For these constraints to
be most useful, they should be based on more than one class of object,
identified in more than one way.  At present, optical astronomy
provides the most sensitive searches for high redshift galaxies. The
two major ways of finding galaxies at high redshift using optical data
are the Lyman Break method (e.g., Steidel et al 1996) and the \lya\ method
(e.g., Hu, Cowie, \& McMahon 1998, Rhoads et al 2000).  

Samples of galaxies selected by these two methods have been compared
in past using broad band spectral energy distributions.  While both
selection methods identify actively star forming galaxies, there are
clear quantitative differences.  The \lya\ selected objects are seen
in most cases to be young, relatively less massive, and less evolved
chemically than Lyman-break selected galaxies (Gawiser et al. 2007,
Pirzkal et al. 2007, Finkelstein et al. 2009, 2010a).  

We now wish to know if the differences between \lya\ and Lyman break
samples are due only to the difference in their average continuum
magnitudes, or if the presence of \lya\ in emission tells us something
significant about the nature of the galaxy.  To address this, we here
perform the first systematic study of the sizes and surface
brightnesses of \lya\ selected galaxy samples over a wide range of
redshift, and compare our result to previous studies of both
size evolution and surface brightness evolution in Lyman break
samples.

Size evolution of the Lyman break galaxy population has been
investigated both for color-selected samples (Ferguson et al. 2004,
Bouwens et al. 2004), and for spectroscopically selected samples
(Hathi, Malhotra, \& Rhoads 2008).  The sizes of typical LBGs increase
from redshift 6 to redshift 1.5.  ``Typical LBGs'' here means those
galaxies with luminosities about $L_*$ or brighter.  The observed size
evolution of LBGs is well described by $R \propto H^{-1}(z)$, as expected for
samples of disk galaxies selected by fixed circular speed.

Despite their smaller characteristic sizes, high redshift LBGs are
similar to low redshift ($z\approx 0$) starbursts in one particular
aspect: their star formation intensity (SFI), measured in luminosity
per unit area, shows a distinct upper limit that does not change from
redshift zero to 6.5 (Hathi et al. 2008, Meurer et al. 1997). The
upper limit of SFI corresponds to the maximum pressure that can be
supported by the interstellar medium (ISM), beyond which galaxy scale
winds become prevelant and inhibit further star formation by heating cold ISM
and/or removing cold ISM with winds (Meurer et al. 1997, Heckman et
al. 1990, Thompson et al. 2005). 

In section 2, we describe the \lya\ samples used in the study, together with
the size measurements.  We present new size measurements for about
half of the galaxies in the overall sample, while we adopt previously
published size measurements for the other half.  In section 3, we
describe our calculations of the surface luminosity and star formation
intensity in the full sample.  We discuss our results in section 4.
Finally in section 5 we summarize the implications of our findings for
galaxy formation and evolution.

\section{Samples and Size Measurements}

We measured the sizes of known Lyman-alpha emitters from several
surveys spanning this range of redshifts (see table~1).  We select
surveys in fields where images from the Hubble Space Telescope (HST)
exist, since \lya\ galaxies are unresolved in ground-based
imaging.\footnote{Here we exclude the Lyman-alpha blobs (e.g. Steidel
  et al 2000, Matsuda et al. 2004, Yang et al 2009, 2010), which come
  in a variety of sizes, appear to be associated with only the most
  overdense regions, and may be powered by some mechanism besides star
  formation.}  We measure the sizes of the \lya\ galaxies from
broad-band imaging around the wavelength 1500\AA, to facilitate
comparison with the sizes measured for LBGs.

The most common measure of size is the half-light radius $r_{hl}$,
defined as radius of a circle enclosing half the total projected light
of the galaxy, and is calculated using the program SExtractor (Bertin
\& Arnouts 1996) in most cases.  The errors in the $r_{hl}$
measurements range from 10\% to 25\% depending on the signal to noise
of the detection (Bond et al. 2009). We measured
the half light radii of \lya\ emitters at $z=2.25$ (a sample from
Nilsson et al 2009), $3.1$ (sample from McLinden et al. 2011), and
$4.45$ (sample from Finkelstein et al 2009), after excluding sources
identified as active galactic nuclei (AGN) (Nilsson et al 2009,
McLinden et al 2010b).  We used the SExtractor half light radius
parameter, based on publically available HST images.  In the $z=2.25$
and $z=3.1$ samples, the HST images are from the COSMOS survey
(Scoville et al 2007; Capak et al 2007), using the
Advanced Camera for Surveys/Wide Field Camera with the F814W filter.
This corresponds to near-UV rest wavelengths of 2000\AA\ at $z=3.1$
and 2500\AA\ at $z=2.25$.  In the $z=4.45$ sample, the HST images are
from the GOODS survey (Giavalisco et al 2004), F775W images provide a
rest wavelength of 1420\AA\ at $z=4.45$. In addition, we compile other
size measurements made using HST images for other Lyman-alpha galaxies
up to redshift $z\sim 6$ (Pirzkal et al. 2007, Bond et al. 2009,
Overzier et al. 2006, 2008, Venemans et al. 2005, Taniguchi et
al. 2009).

In table 1, we summarize the different samples for which we have
measured the sizes of the Lyman-alpha galaxies, including the number
of objects and redshift for each sample, and the continuum rest
wavelength used for size meaurements.  The measured sizes of galaxies
are not very sensitive to the wavelength used, given the modest
wavelength differences among the samples we use (Ferguson et
al. 2004). In all, our samples
comprise 368 galaxies, among which we are reporting the first size
measurements for 171.

About 15-30\% of \lya\ galaxies are also reported to be ``clumpy'',
where the size of individual clumps is small but the separation
between the clumps is large, $\approx 10-20$ kpc (Pirzkal et al. 2007,
Rhoads et al. 2005, Taniguchi et al. 2009, Bond et al. 2009).  These
galaxies are included in the analysis, taking the interclump distance
as the size. Being a minority, they do not affect the average size
greatly.

\section{Surface Luminosity Measurements}
We calculate the surface luminosity following the method described 
by Hathi et al. 2008, for deriving dust corrected UV flux at rest-frame
2300 \AA.  We assume that the UV continuum is well represented by a 
power law of the form $ f_{\lambda} \propto \lambda^{\beta} $, 
where  $f_{\lambda}$ is  the flux  density per  unit  wavelength (ergs
s$^{-1}$ cm$^{-2}$ \AA$^{-1}$).

The UV spectral slope ($\beta$) is then derived from a power
law fit to the UV colors using two filters.  In cases where we did not
have the color information, we assume that the UV slope $\beta = -2$,
which is a typical slope among well studied \lya\ galaxies  (Pirzkal
et al. 2007, Gawiser et al. 2007, Nilsson et al. 2009, Rhoads et
al. 2009).

A dust free stellar population is expected to have a slope of $\beta
\approx -2$ for a wide range of young stellar populations, so observed
deviations from this slope give us an estimate of reddening by dust.  This
reddening estimate is then used along with the Calzetti extinction law
(Calzetti, Kinney, \& Storchi-Bergmann 1994) to
correct for dust extinction.  We next k-correct the UV flux at the
observed wavelength to that at rest-frame 2300\AA.  Finally, we apply
a correction factor to convert the extinction corrected UV luminosity
into a bolometric luminosity: ${L_{UV}} / {L_{\rm bol}} \simeq 0.33$.
We have checked this ratio against detailed stellar population
synthesis models (Bruzual \& Charlot 2003), and find good agreement,
with the models giving $0.2 < {L_{UV}} / {L_{\rm bol}} < 0.5$ (Hathi
et al. 2008).

The Star-Formation Intensity $S$ is then defined using the luminosity and
size.
\[
{\rm S} = \frac{L_{\rm bol}}{2 \pi r_e^2}  \; \; \;  \left (\frac{L_{\sun}}{{\rm kpc}^2} \right ) ~~.
\]      
The SFI vs redshift is plotted in Figure 1, for LBGs (Hathi et
al. 2008) as well as \lya\ emitters.

\begin{deluxetable}{lccccll}
\tablehead{
\colhead{z}    &  \colhead{rest wave-}   & \colhead{sample} & \colhead{mean $R_{hl}$ }  &
\colhead{$\sigma(R_{hl})$} & \colhead{Sample} & \colhead{Size}\\
\colhead{}    &  \colhead{length (\AA)}   & \colhead{size} & \colhead{}  &
\colhead{} & \colhead{reference} & \colhead{reference}\\
}
\startdata
2.25 & 2500 & 142 & 0.15 & 0.04  & Nilsson et al. 2009 &This work\\
3.13 & 2000 & 13  & 0.11 & 0.03  & Venemans et al. 2005 &Venemans et al 2005\\
3.12 & 2000 & 15  & 0.14 & 0.06  & McLinden et al. 2010 &This work\\
3.1  & ...  & 110 & 0.145 & 0.07 & Bond et al. 2009 &Bond et al 2009\\
4.1  & 1670 & 12  & 0.135 & 0.05 & Overzier et al. 2008 &Overzier et al 2008\\
4.45 & 1420 & 14  & 0.12  & 0.02 & Finkelstein et al. 2009 &This work\\
5.2  & 1370 & 4   & 0.125 & 0.03 & Overzier et al. 2006 &Overzier et al 2006\\
5.7  & 1220 & 49  & 0.16 & 0.06  & Taniguchi et al. 2009 &Taniguchi et al 2009\\
4--5.7 & ... & 9   & 0.16 & 0.08  & Pirzkal et al. 2007 & Pirzkal et al 2007
\enddata
\end{deluxetable}

\section{Results}

{\bf 1. Star-formation Intensity} Hathi et al. (2008) showed that the
upper limit for star formation intensity (SFI) seems to hold for
Lyman-break Galaxies up to redshifts $z > 6.5$.  In figure 1 we see
that the SFI for the \lya\ emitters also stays within that upper
envelope.  The lower envelope of the SFI distribution is of course set
by incompleteness, so we concentrate on the upper envelope in our
studies.

The typical SFI for \lya\ galaxies is smaller than that of LBGs by
about a factor of 2.5.  This is still above the threshold required to
drive galaxy-scale winds (Lehnert \& Heckman 1996; Heckman 2001), 
and such winds may play a
significant role in \lya\ photon escape (e.g. Verhamme, Schaerer, \&
Maselli 2006; Steidel et al 2010). A Kolmogorov-Smirnov test shows
that the SFI is different at 5-$\sigma$ level between \lya\ and Lyman
break samples.

The reason for the smaller SFI in \lya\ galaxies compared to LBGs is
not immediately clear. It could be due to an overestimate of the sizes
of \lya\ galaxies, which are only marginally resolved with HST. An
overestimate of sizes by 50\% can largely explain the discrepancy
between the surface luminosity of \lya\ and LBGs.  This effect should,
however, be important for $z > 5$ LBGs as well, since they are also
barely resolved by HST.

\begin{figure}[h]
\plotone{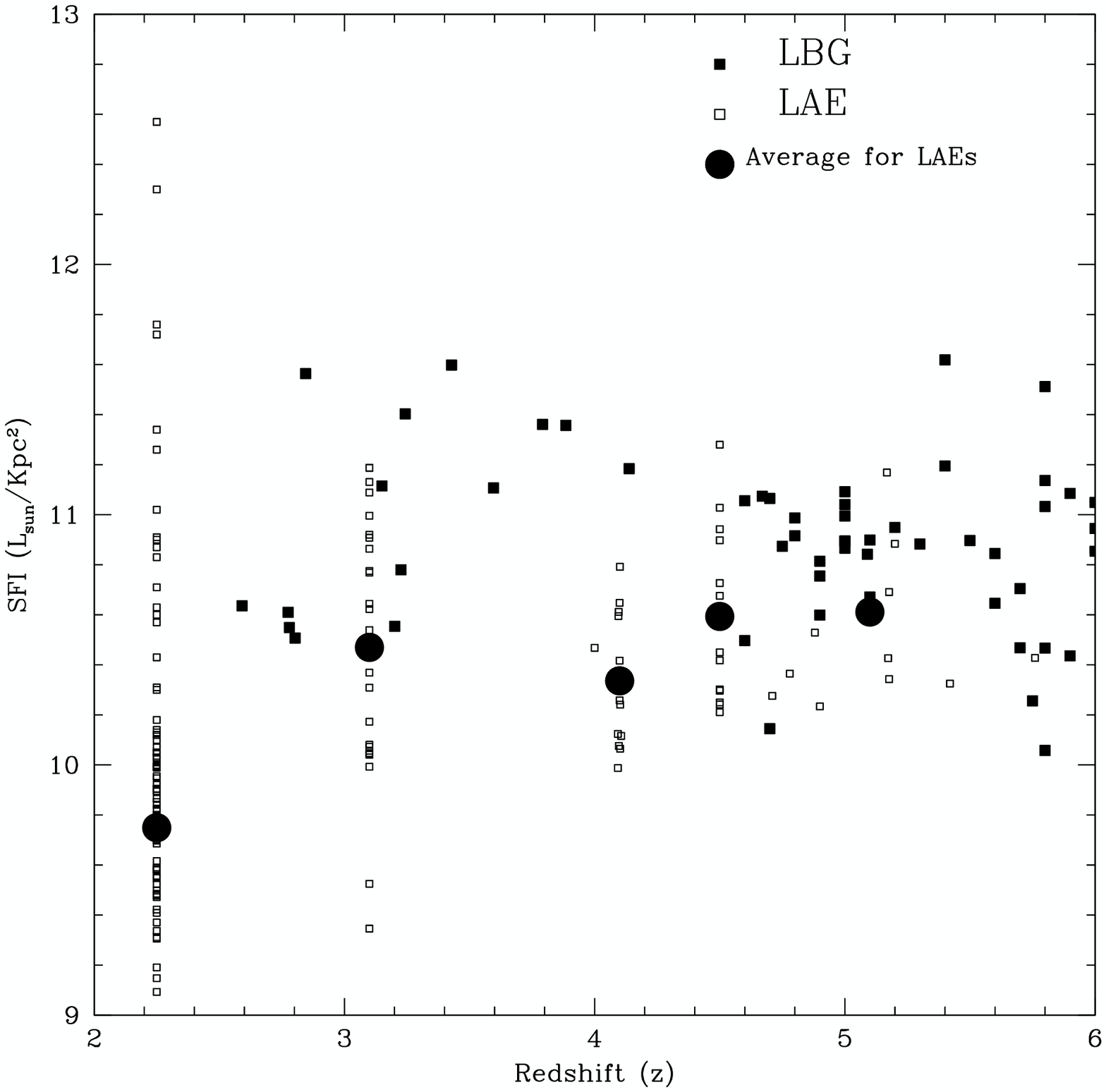}
\caption{The Surface Luminosity (or Star Formation Intensity) is
  plotted aginst redshift for Lyman Break Galaxies (solid squares) and
  Lyman-$\alpha$ galaxies (open squares), with the averages for the
  \lya\ galaxies shown as large filled circles. The SFI for \lya\
  galaxies is generally lower than that of LBGs, but shows no
  evolution with redshift.}
\end{figure}

{\bf 2. Sizes}
Figure 2 shows the sizes of \lya\ and Lyman-Break galaxies, as
represented by half-light radii.  The size of both the \lya\ galaxies
and LBGs are measured in the UV continuum, and not in the line
emission.  We see from Figure 2 that the median size of \lya\ galaxies
is small and does not change with redshift $z$ over the range 
$2.25 < z < 6.6$. 
Given that nearly all of the \lya\ galaxies studied here were 
identified in ground-based surveys with typical seeing of $\ga 1''$, 
it seems unlikely that the typical observed size of $\sim 0.15''$ is 
due to observational selection against larger objects.
In contrast, the LBGs at lower redshifts are larger than the LBGs at high
redshifts (Ferguson et al. 2004, Bouwens et al. 2004, Hathi et
al. 2008). The average size of LBGs follows closely the relation 
$r_{hl} \propto 1/H(z)$, where $H(z)$ is the Hubble parameter at
redshift z. This is not to say that it is the same set of \lya\
galaxies that stay unevolved from redshift ~2 to ~6.  It just means
that whenever we identify galaxies by means of strong \lya\ emission,
we find them to be compact in size, independent of the redshift. It is
very likely that \lya\ emitting phase in a galaxy is short with a duty
cycle of about 10\% -- 15\% (e.g. Malhotra \& Rhoads 2002, 
Kova\v{c} et al. 2007, Tilvi et al. 2009, Nagamine et al 2010).

\begin{figure}[h]
\plotone{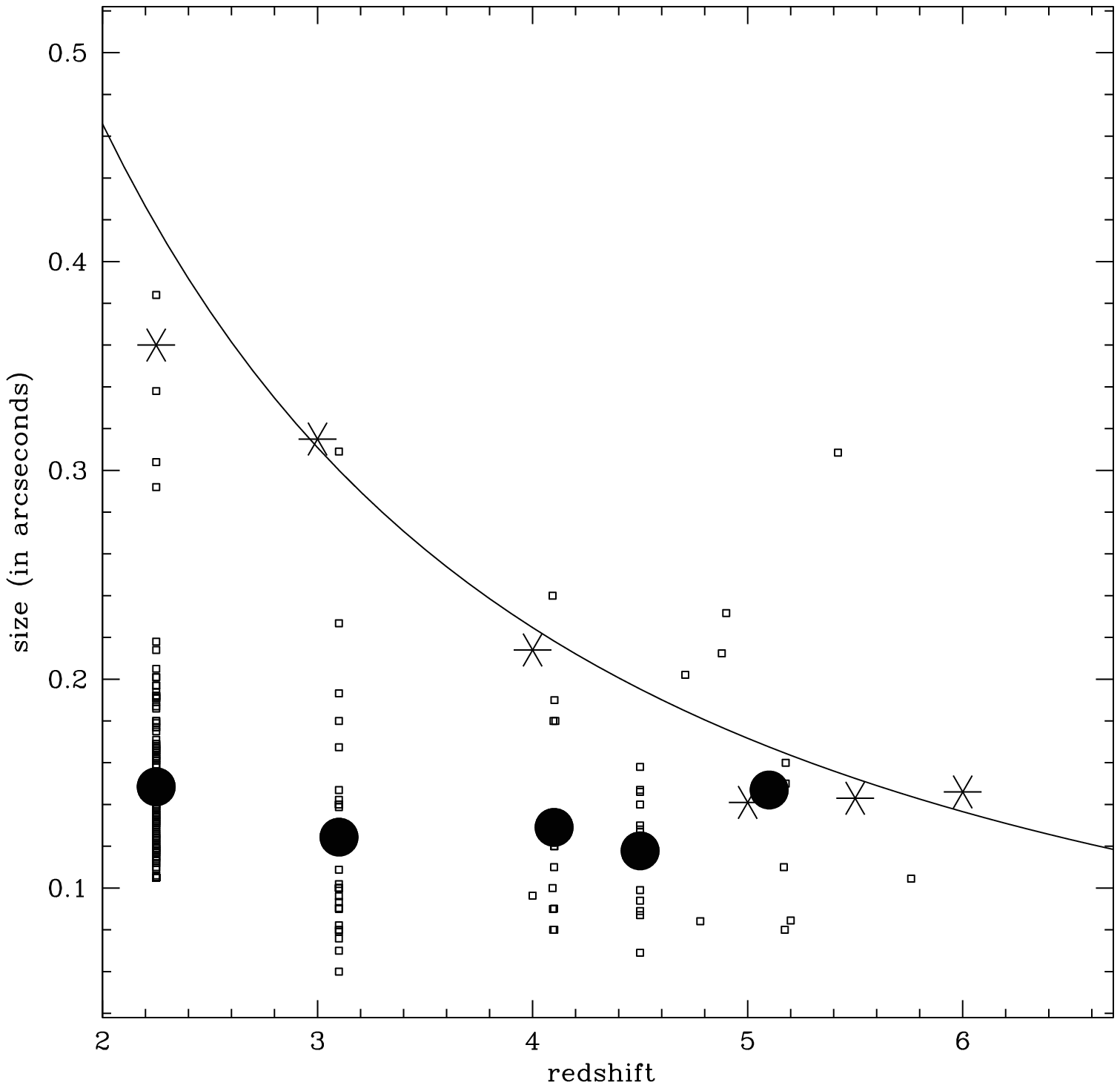}
\caption{ Angular sizes of \lya\ galaxies (open squares) are plotted
  against redshift, with the averages given by dark circles. The stars
  show mean angular sizes of Lyman Break Galaxies. While LBG sizes
  increase as we go to lower redshift (and an older snapshot of the
  universe), the \lya\ galaxies are always seen to have the same size. At
  redshifts greater than 5, both \lya\ and LBGs show similar sizes, and
  have similar physical properties. The samples plotted here are listed
  in Table 1.}
\end{figure}

{\bf 3. Luminosity} Putting together the evolution in size and the
constancy of surface luminosity for both the LBGs and LAEs, we can
reproduce the observed evolution of $L_*$ in LBGs and LAEs. The UV
luminosity function of \lya\ galaxies does not evolve measurably for
redshifts $3 <z < 6.0~~$ (Ouchi et al. 2008). On the other hand the
luminosity function of LBGs does evolve, with $L_*$ getting brighter
at lower redshifts (e.g. Bouwens et al. 2008).  In figure~3, we plot
our prediction for the redshift evolution of $L_*$--- based on the
observed size evolution and constant characteristic surface brightness
of LBGs--- and plot for comparison the observed evolution in $L_*$.
We see that the observed evolution is reproduced well by our model,
except for the lowest redshift points, where corrections for dust
extinction become important.  In deriving the constant upper envelope 
to star formation intensity, both Meurer et al (1997) and Hathi et al
(2008) applied dust extinction corrections based upon the observed spectral
slope $\beta$.  However, the characteristic luminosity $L_*$ and 
corresponding magnitude $M_*$, as plotted in figure~3, are uncorrected 
for dust.  At $z=5$, the measured $M_*$ falls on the model curve. 
At $z=4$, Hathi et al (2008) report a value of $\beta$ that is
redder by $0.1$ in power law slope, corresponding to an $0.2$ mag
increase in $1600$\AA\ dust extinction, while the observed $M_*$ is
$0.4$ mag fainter than the model.  At $z=3$, the $\beta$ is redder
by 0.5 units, the predicted extinction becomes 1 magnitude (relative
to $z=5$), and the observed offset is about $0.8$ mag.  Thus,
within the uncertainties, our treatment provides a good explanation 
for the evolution of $M_*$, and dust extinction effects can account
for the most significant deviations between the model and the observations.

A full calculation of the luminosity function would depend on the
bivariate probability distribution of galaxy size and galaxy surface
brightness.  However, by restricting our attention to $L_*$, we avoid
much of this complexity.  Since $L_*$ is a ``knee'' in the luminosity
function, above which the number of galaxies drops exponentially, we
expect that $L_*$ galaxies will correspond roughly to the upper
envelope of star formation intensity.

\begin{figure}[h]
\plotone{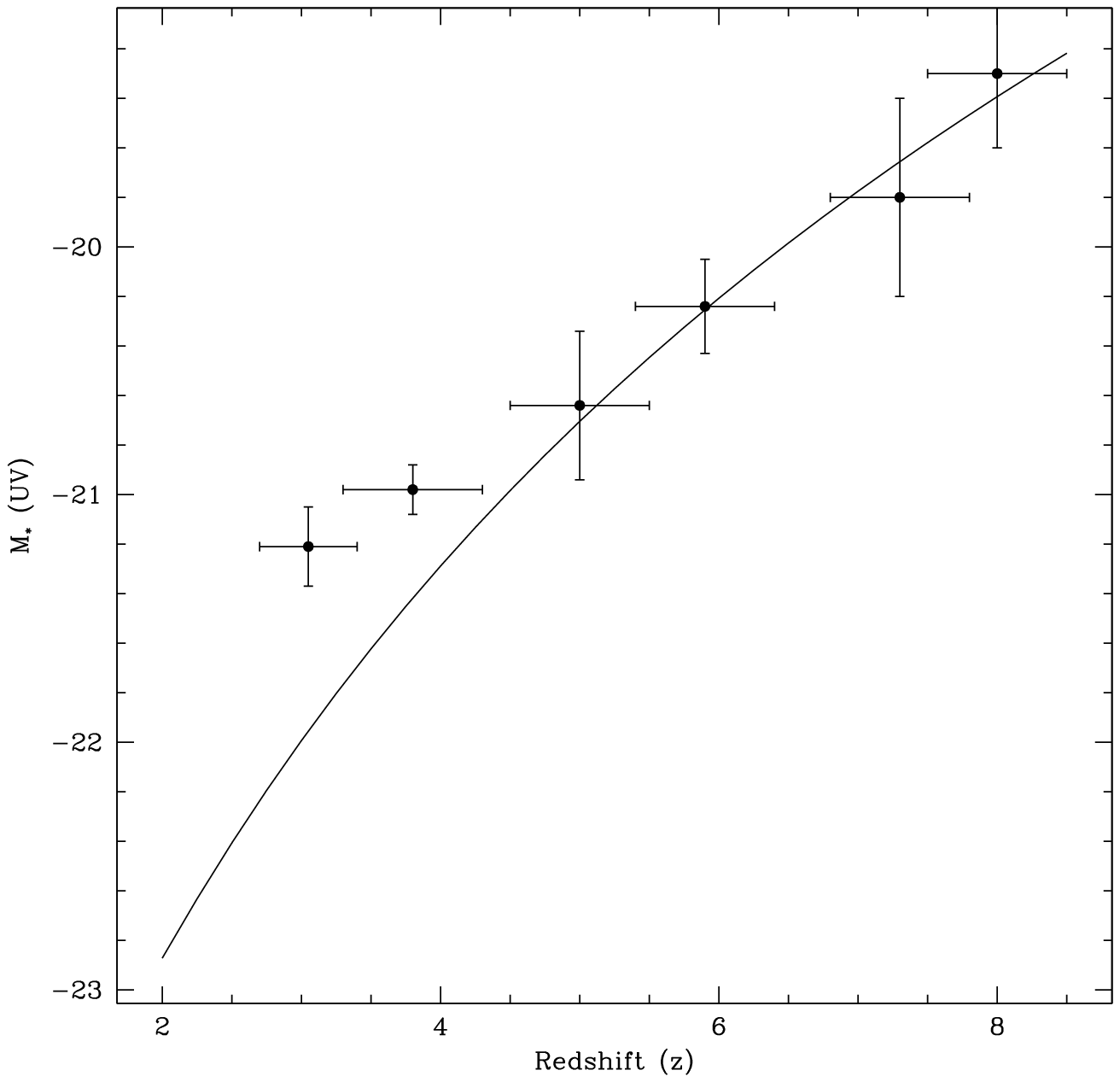}
\caption{ By combining the size of LBGs with upper limit to luminosity
  intensity, we can predict the redshift evolution of $L_*$, the
  characteristic luminosity of Galaxies. This plot shows the redshift
  dependence of the absolute magnitude $M_*$, which is related to the
  characteristic UV luminosity $L_*(UV)$ by $L_*(UV) \propto 10^{-0.4
    M_*}$.  This model is represented by the line. The points
  represent the observed absolute magnitudes $M_*$ of Lyman Break
  Galaxies at different redshift bins (Bouwens et al 2007). 
  At lower redshifts, the
  extinction correction can explain the discrepancy between the model
  and the observed points (see text). }
\end{figure}

\section{Discussion}

At this point the alert reader will note that of the three quantities
radius, luminosity, and surface luminosity, only two are
independent. So once we have determined the size evolution with
redshift and know that the upper envelope of the surface brightness
stays constant, we can then predict the redshift evolution of
$L_*$ (fig.~3). The relevant question then becomes: of these three properties,
which two are the independent variables, and which one is the
dependent property?  For that, we will rely on which two quantities
seem more physically fundamental.

The upper limit to the surface luminosity is a natural candidate. The
upper limit for LBGs and low-redshift starbursts observed by Meurer et
al 1997, Hathi et al. 2008, and here, corresponds to the maximum
pressure measured in starburst galaxies driving winds (Heckman, Armus
\& Miley (1990). The LBGs also show evidence for strong winds in the
line profiles and relative velocities between absorption and emission
lines (e.g. Steidel et al. 2010).

The observed size evolution of LBGs can be understood in terms of
growth of disks.  If LBG selection identifies galaxies with a
particular rotation speed at all redshifts, we should expect their
typical size to grow with redshift as $H(z)^{-1}$ (Fall \& Efstathiou
1980).  Of course, it is not entirely clear that the LBGs, or \lya\
galaxies, are disks in formation. Ravindranath et al. 2006 find that
only 40\% of LBGs at $z \approx 3$ have exponential profiles. Similar
results are seen for \lya\ galaxies at z=3.1 (Gronwall et al. 2010).

% The size evolution of LBGs can be understood in terms of growth of
% disks. Keeping the rotational velocity constant leads to disk sizes to
% grow as $H(z)^{-1}$ (Fall \& Efstathiou 1980)as seen for LBGs
% (Ferguson et al. 2004), whereas keeping the mass constant gives
% $H(z)^{-2/3}$ dependence.  It is not yet clear that the LBGs, or \lya\
% galaxies, are disks in formation. Ravindranath et al. 2006 find that
% only 40\% of LBGs at $z \approx 3$ have exponential profiles. Similar
% results are seen for \lya\ galaxies at z=3.1 (Gronwall et
% al. 2010).

It could well be that \lya\ galaxies show us a very early stage of
galaxy formation, as initially suggested by Partridge \& Peebles
(1967). This is supported by the small physical sizes, smaller dust
extinction, younger stellar ages and lower stellar masses seen for
most (though not all) \lya\ galaxies (Pirzkal et al. 2007, Gawiser et
a. 2007, Finkelstein et al. 2007, 2008, 2009, Pentricci et
al. 2008). The first observational demonstration of the extreme youth
of the stellar populations in these galaxies came from the high
equivalent widths of the \lya\ line, which arises most naturally from
stellar populations younger than 30 Myr (Malhotra \& Rhoads
2002). Studies of correlation functions of \lya\ galaxies also
indicate that the duty cycle of \lya\ emitting phase is about 15\%
(e.g., Kova\v{c} et al 2007).  The bias inferred from the correlation
function matches that of low redshift $L_*$ galaxies, further
supporting the idea that \lya\ emitters may be building blocks of
typical present-day galaxies (e.g. Gawiser et al 2007).  On the other
hand, there is some controversy whether those Lyman break-selected
galaxies showing \lya\ line emission are older or younger than their
counterparts without line emission (Kornei et al. 2010, Pentericci et
al 2008). Perhaps size is the crucial property that determines whether
or not a galaxy shows \lya\ emission.

So far, we have discussed the sizes of galaxies as seen in the UV
continuum. This allows us to compare the LBGs and \lya\ galaxies on an
equal basis.  The \lya\ galaxies may show a different size in line
emission: larger, if the resonant scattering of \lya\ photons to
escape from the galaxy is important; and smaller if the \lya\ comes
from an active nucleus.  Rhoads et al (2009) examined the sizes of
galaxies in the \lya\ line and the adjoining continuum using data from
slitless spectroscopy, and found no significant difference in FWHM in
the line and continuum.  HST imaging in narrow bandpass dominated by
the \lya\ line flux also exists for a few cases (Bond et al. 2010,
Finkelstein et al. 2010b). Bond et al (2010) found no measurable
difference in sizes using narrow and broad-band images.  Finkelstein
et al. (2010b) examined three z=4.4 LAEs in HST narrowband imaging and
found them to belarger in \lya\ than in their rest-frame UV continuum
light, with r$_{h}$ = 1.20 $\pm$ 0.03 and r$_{h}$ = 0.80 $\pm$ 0.02
kpc for the \lya\ and UV light, respectively.  This implies that
resonant scattering plays a role in LAEs, as it is altering the escape
path of \lya\ photons.  A similar result has recently been obtained by
stacking LBGs, which shows that extended low surface brightness \lya\
emission may be a common feature of LBGs (Steidel et la 2011).

\section{Conclusions}

We have shown that the growth in characteristic luminosity of high
redshift star-forming galaxies proceeds in step with their growth in
size, due to their observed maximum surface brightness (which in turn
may be explained by galactic wind feedback).

We have also shown that a compact physical size is a key factor in
whether or not a galaxy is a \lya\ emitter.  This is true for all
redshifts in the range $2\la z \la 6$, and \lya\ galaxies are faint at these
redshifts because of their small sizes. The more fundamental question,
then, is why do we see the same size for galaxies with \lya\ emission
across the redshift range 2--7? The \lya\ line is resonantly
scattered, which increases its path length in neutral gas and
therefore the chance of its being absorbed by dust--- although in a
clumpy medium it can escape (Neufeld 1990, Finkelstein et
al. 2008, 2009). Galactic winds also help in the escape of \lya\ photons by
shifting the photons away from the resonant frequency (Santos 2004, 
Verhamme et al 2006, McLinden et al. 2010).  
It seems likely that larger galaxies have larger optical
depths, so that \lya\ photons are resonantly scattered and we no
longer see the line in emission. It could also be that larger
galaxies contain multiple starbursting regions, resulting in
incoherent flows on a galaxy-wide scale.  This in turn could hamper
\lya\ escape, since any particular line of sight would be likely to
sample gas at a wide range of velocities.
Larger, more massive galaxies also
tend to have more metals from previous generations of star-formation,
leading to more dust and fewer \lya\ photons escaping.  While we do
not completely understand the relative importance of all the
mechanisms that aid in \lya\ escape, the small sizes of these galaxies
give us a crucial clue to their nature.

\acknowledgments
We are grateful to Evan Scannapieco, Gerhardt Meurer and Mike Fall for useful discussions. We also acknowledge financial support from the US National Science Foundation through NSF grant AST-0808165.  We thank the DARK Cosmology Centre in Copenhagen, Denmark, for hospitality during the completion of this work.

\def\apj{{\it Astrophys. J.}}
\def\apjsupp{{\it Astrophys. J. Supp. }}
\def\aj{{\it Astron. J.}}
\def\aap{{\it Astron. Astrophys.}}
\def\aapsup{{\it Astron. Astrophys. Supp.}}
\def\mnras{{\it Mon. Not. R. Astron. Soc.}}

\end{document}